# First-Principles Study of Transition Metal Doped in 2D Polyaramid for Novel Material Modelling

Ravi Trivedi[1,2], Chaithanya Purushottam Bhat[3], Shakti S. Ray[1, 2], Debashis Bandyopadhyay[3*]

[1]Department of Physics, Karpagam Academy of Higher Education, Coimbatore 641021, Tamil Nadu, India

[2]Centre for Computational Physics, Karpagam Academy of Higher Education, Coimbatore 641021, Tamil Nadu, India

[3*]Department of Physics, Birla Institute of Technology and Science, Pilani,

Rajasthan - 333031, India, e-mail: debashis.bandy@gmail.com

**Abstract**

We present a first-principles density functional theory (DFT) study of transition metal (TM = Ti, Cr, Mn, Fe, Co, Ni) functionalized two-dimensional polyaramid (2DPA) to explore their structural, electronic, and magnetic properties. Mechanical parameters, such as bulk modulus, shear modulus, Young's modulus, Poisson's ratio, and Pugh ratio, together with phonon dispersion, confirm the mechanical and dynamic stability of all doped systems. Electronic structure analysis shows strong binding of Co, Cr, Fe, Ni, and Ti with formation energies between –1.15 eV and –2.96 eV, while Mn binds more weakly (-0.67 eV). TM doping introduces new electronic states that reduce the band gap, with Fe-doped 2DPA exhibiting the lowest value of 0.26 eV. The systems display predominantly ferromagnetic ordering, with magnetic moments of 1.14 μB (Co), 3.57 μB (Cr), 2.26 μB (Fe), 4.19 μB (Mn), and 1.62 μB (Ti). These results demonstrate that TM-doped 2DPA possesses tunable magnetic and electronic characteristics, highlighting its potential for spintronic applications.

**Graphical**

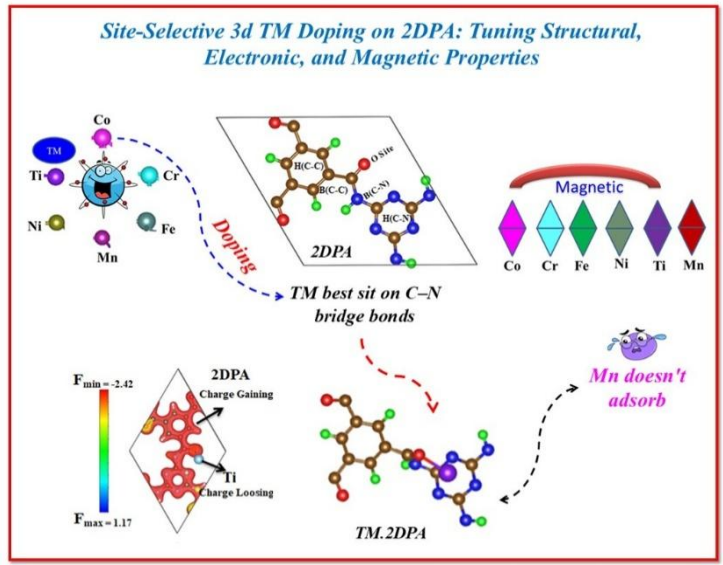

## 1. Introduction

In recent years, two-dimensional (2D) materials have garnered substantial interest due to their exceptional optical, mechanical, and electrical properties, which markedly differ from their bulk counterparts. Materials like graphene, transition metal dichalcogenides (TMDs), phosphorene, and MXenes are particularly notable for their high surface area-to-volume ratio and reduced dimensionality, making them highly suitable for advanced applications such as nanoelectronics, catalysis, and energy storage [1–5].One of the most effective strategies for tuning the properties of 2D materials is doping with transition metals (TMs). TM doping introduces new electronic states and modifies structural characteristics, enabling functionalities that are otherwise unattainable in pristine materials [6–10]. This has opened new directions in spintronics, magnetic storage, and electronics, where tuning electronic and magnetic properties is crucial. For instance, doping semiconducting 2D systems like $MoS_2$ and $WS_2$ with TMs such as Re, Mn, or Fe has been shown to induce localized states, adjust band gaps, and introduce magnetic ordering—key properties for transistors, sensors, and spintronic devices [11–13]. Magnetism can also be induced in typically non-magnetic materials. For example, TM doping in graphene (e.g., with Fe, Co, or Ni) introduces magnetic moments and modifies the band structure [14]. Similarly, doping TMDs with V or Mn results in magnetic ordering, enhancing their suitability for spin-based technologies [14,15]. The ability to tune magnetic properties at the atomic scale is particularly significant for designing magnetic sensors, memory devices, and other spintronic components [15–17]. In this context, two-dimensional polyaramid (2DPA) has emerged as a promising candidate for magnetic functionalization. Synthesized via solution-phase polymerization, 2DPA distinguishes itself with its porous architecture and high specific surface area, attributed to large interconnected benzene and triazine rings linked by amide bonds [18,19]. With a semiconducting band gap of ~2.452 eV and a carbon-based backbone enriched with nitrogen and oxygen atoms, 2DPA resembles other functional materials like covalent triazine frameworks and graphitic carbon nitrides in terms of energy-related applications [20,21].

Despite substantial studies on TM doping in materials such as graphene and TMDs, the specific effects of 3d transition metals (Co, Cr, Fe, Mn, Ni, Ti) on the surface properties of 2DPA remain insufficiently explored. Given 2DPA's distinctive structure, it provides an ideal platform for investigating how TM dopants interact at the atomic level, influence magnetic behavior, and impact the electronic structure. These modifications are critical for extending its use in spintronic devices, catalysis, and energy storage applications.

To date, a systematic theoretical investigation of single-atom TM doping in 2DPA has not been reported. This study seeks to bridge that gap using first-principles density functional theory (DFT), a widely accepted computational method for evaluating the electronic and magnetic behavior of complex systems [22–37]. DFT has proven effective in modeling TM doping effects in other 2D materials and is thus well-suited for analyzing the induced magnetic moments and potential magnetic ordering in TM-doped 2DPA.

Through detailed DFT analysis, we aim to elucidate the interplay between dopant type, surface interaction, and magnetic response in 2DPA. The findings will offer new insights into surface magnetism in 2D polymers and inform future applications of 2DPA in next-generation spintronic and multifunctional electronic devices

## 2. Methodology

In this study, we performed first-principles density functional theory (DFT) calculations to explore the structural, electronic, and magnetic properties of two-dimensional (2D) polyaramid (2DPA) materials doped with 3d transition metals (TMs), including Ti, Cr, Mn, Fe, Co, and Ni. All calculations were conducted using the Vienna Ab initio Simulation Package (VASP) [38–40], employing the projector augmented wave (PAW) method to describe the electron-ion interactions [41,42]. Exchange-correlation effects were treated within the generalized gradient approximation (GGA) using the Perdew-Burke-Ernzerhof (PBE) functional [43]. While the GGA-PBE functional is efficient and widely used, it often underestimates the band gap of semiconducting and insulating systems. To improve the accuracy of electronic correlation effects, especially for localized d electrons in transition metal dopants, we employed the GGA+$U_{eff}$ approach [44]. The effective Hubbard U ($U_{eff}$) values were chosen based on prior studies: 5.0 eV for Co, 3.5 eV for Cr, 4.0 eV for Mn, 4.6 eV for Fe, 5.0 eV for Ni, and 4.4 eV for Ti. The pristine 2DPA structure was fully optimized to obtain its ground-state geometry. A plane-wave energy cutoff of 500 eV was used throughout. The Brillouin zone was sampled using a Monkhorst-Pack k-point mesh of 7×7×1 for structural relaxation and 9×9×1 for density of states (DOS) calculations [45]. For band structure computations, non-self-consistent field (NSCF) calculations were carried out along high-symmetry paths of the Brillouin zone. A vacuum spacing of 25 Å was introduced along the z-direction to avoid interlayer interactions in the periodic slab model. To investigate the preferred doping configuration, five potential adsorption sites for TM atoms were considered: bridge C–N, bridge C–C, hexa C–C, hexa C–N, and edge O positions (as illustrated in Figure 1). Each configuration was optimized until the Hellmann–Feynman forces on all atoms were less than 0.01 eV/Å, and electronic convergence was set to $10^{-6}$ eV.

Mechanical stability was assessed using elastic constant calculations via the USPEX toolkit and the method proposed by Mazhnik and Oganov [46,47]. Dynamical stability was further evaluated through phonon dispersion calculations using the PHONOPY code [48].

Electronic properties of pristine and TM-doped 2DPA systems were analyzed through DOS and band structure calculations. The pristine material was found to be a direct band gap semiconductor. Upon TM doping, a reduction in band gap was observed, accompanied by modifications in the Fermi level position, suggesting doping-induced transitions from n-type to p-type semiconducting behavior. These changes in the electronic structure are essential for tailoring the material's functionalities for applications in nanoelectronics, spintronics, and sensor technologies.

## 3. Results and discussion
### 3.1 Structural Analysis

Two-dimensional polyaramid (2DPA), derived from polymeric polyaramid chains, is a structurally robust material with high thermal stability and mechanical strength. Its structure comprises repeating aromatic rings linked via amide (–CONH–) groups, which form a planar, sheet-like configuration upon reduction to two dimensions. Structural optimization yielded lattice parameters of a = b = 11.13 Å (Fig. 1), confirming the expected hexagonal periodicity.

Detailed atomic coordinates and bond lengths are provided in the Electronic Supplementary Information (SI). Key bond distances include the average bond length for the carbon-carbon (C−C) bonds within the hexagonal rings ($C_h$−$C_h$) is 1.402 Å, while the distance between the hexagonal ring carbon ($C_h$) and the amide carbon ($C_b$) is 1.512 Å. The bond lengths for the carbonyl (C=O) and amide (C−N) bonds are 1.226 Å and 1.392 Å, respectively, with the C−N bonds in the triazine ring ($C_t$−$N_t$) measuring an average of 1.342 Å. These values are consistent with earlier theoretical and experimental reports on related carbon-based polymeric and triazine systems [49–51], validating the structural fidelity of the 2DPA surface. To explore the tunability of 2DPA for potential applications, we investigated its surface doping behavior with 3d transition metals (TMs). Owing to its porous and symmetric hexagonal structure (SI, Fig. 1), pristine 2DPA offers several potential adsorption sites. Five high-symmetry doping positions were examined: bridge C−N, bridge C−C, hexa C−C, hexa C−N, and edge O positions (Fig. 1), in line with prior investigations on nitrogen-rich carbon materials such as nitrogen-enriched polytriazine (NPT) [52]. Our findings show that the site preference is dopant-dependent. Co, Cr, and Fe tend to favor adsorption at the hexa C−C and hexa C−N sites, forming strong interactions with the 2DPA carbon framework. In contrast, Mn and Ti exhibit a preference for edge O positions, with weaker surface binding. These differences significantly impact the electronic and magnetic behaviors of the doped systems. Doping alters the electronic structure of 2DPA by introducing localized impurity states that interact with the delocalized π-system of the host lattice. Notably, Fe doping causes a pronounced reduction in the band gap—from 2.452 eV in pristine 2DPA to 0.26 eV in the Fe-doped case—due to hybridization between Fe 3d orbitals and 2DPA π* states [53]. Such gap engineering through doping enhances the electronic conductivity and paves the way for tailoring carrier type and concentration in semiconducting devices. Magnetically, transition metal doping induces substantial local magnetic moments and gives rise to ferromagnetic ordering through indirect exchange interactions between TM d-electrons and the conjugated π-electrons of the host lattice. The magnitude of the induced magnetic moment depends strongly on the dopant: Co (1.14 μB), Cr (3.57 μB), Fe (2.26 μB), Mn (4.19 μB), and Ti (1.62 μB) per unit cell. These values suggest that TM-doped 2DPA can exhibit diverse magnetic characteristics, with potential applications in spin-based electronic devices. Moreover, the dopants influence surface charge density and reactivity. Redistribution of charge due to metal–surface interaction enhances surface polarization, potentially improving catalytic activity and enabling charge-controlled functionalities. The coexistence of localized magnetic moments and modified electronic states further highlights the versatility of TM-doped 2DPA in multifunctional device applications such as spintronic memory, magnetic sensing, and electrocatalysis.

The present study reveals that transition metal doping is an effective strategy to modulate the surface electronic and magnetic properties of 2DPA. The choice of dopant and adsorption site critically determines the extent of band gap narrowing, magnetic ordering, and surface reactivity. These insights establish TM-doped 2DPA as a promising platform for next-generation nanoelectronic and spintronic technologies requiring precise control over surface magnetism and electronic conduction.

To identify the most energetically favorable doping configuration, we calculated the formation energy ($E_F$) for each transition metal (TM) atom using the expression:

$$E_F = E_{2DPA.TM} - E_{2DPA} - E_{TM} \ \ - - - - (1)$$

Where, $E_{2DPA.TM}$ is the total energy of the TM-doped 2D polyaramid structure, $E_{2DPA}$ is the energy of the pristine 2DPA, and $E_{TM}$ is the energy of an isolated TM atom in vacuum.

After structural optimization, the preferred adsorption sites were determined for each TM atom. Transition metals such as Co, Cr, and Fe showed a strong preference for hexa C−C and hexa C−N positions. These sites facilitate robust metal–surface interactions due to their symmetric environment and the proximity of conjugated π-electron systems. In contrast, Ti and Mn exhibited a preference for the edge O site. Although Ti formed a stable configuration at this position, Mn demonstrated significantly weaker interaction, leading to an off-plane geometry due to incomplete incorporation into the 2DPA lattice.

The formation energies reflect the thermodynamic stability of each doped configuration. The calculated $E_F$ values are –1.28 eV for Co, –1.15 eV for Cr, –1.38 eV for Fe, –1.53 eV for Ni, and –2.96 eV for Ti (Table 1). Among these, Ti exhibits the most negative formation energy, indicating the strongest binding and highest structural stability upon doping. These negative values confirm the favorable incorporation of these TMs into the 2DPA framework and suggest strong chemisorption, likely arising from d-orbital hybridization with the π-electrons of the polyaramid surface [53]. In contrast, Mn exhibits a comparatively weak interaction, with a formation energy of only –0.67 eV. This significantly less negative value correlates with its inability to fully integrate into the planar lattice, as visualized in the side-view geometry shown in Fig. 1b. The dopant remains elevated above the 2DPA surface, lacking the orbital overlap necessary for strong binding. This is consistent with its less favorable d-electron configuration, which reduces the degree of hybridization with the host lattice and limits structural and electronic modification. The variation in formation energies among the dopants highlights the role of electronic structure—particularly d-orbital occupancy—in determining dopant-surface interactions. Transition metals like Fe, Co, and Ni, which have partially filled d orbitals, can effectively hybridize with the delocalized π-system of 2DPA. This not only stabilizes the planar geometry but also significantly alters the material's electronic and magnetic properties. Ti, with its highly reactive 3d orbitals, forms the most stable configuration and is thus a particularly promising candidate for further studies in electronic and spintronic applications.

These results emphasize the critical influence of TM selection on the surface behavior of 2DPA. By tuning the nature and position of the dopant, one can engineer the material's surface characteristics, such as electronic structure, magnetism, and catalytic reactivity. The contrasting behavior of Mn also underscores the importance of considering both chemical compatibility and orbital interaction when designing functional materials. Ultimately, these insights provide a pathway toward the rational design of 2DPA-based systems for advanced technologies, including magnetic sensors, data storage, catalysis, and next-generation spintronic devices.

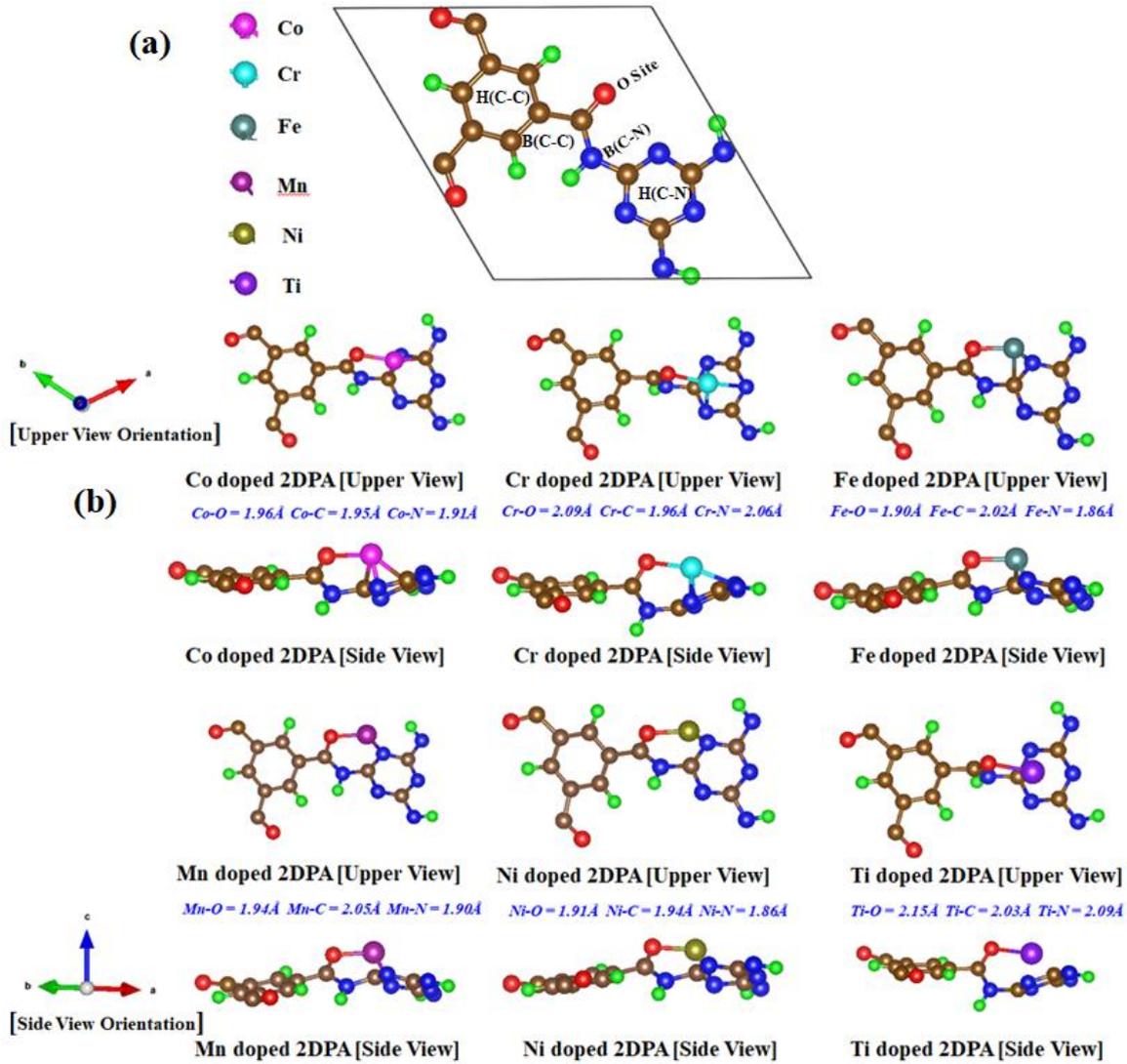

**Figure 1** Schematic representation of (a) unit cell of 2DPA with available adsorption sites at five locations where H, B indicate hexa and bridge sites (b) minimum energy structure of TM (Co, Cr, Fe, Mn, Ni, and Ti) doped 2DPA with upper and side view orientation. [Blue, brown, red, and green ball shows Nitrogen, Carbon, Oxygen, and Hydrogen atoms, respectively]

**Table 1** Adsorption energy ($E_{ads}$), Formation distance of TM over the bridge C-N of TM-doped 2DPA.

| TM doped 2DPA | $E_{ads}$ (eV) | Formation distance (Å) |
|---|---|---|
| Co | -1.28 | 1.94 |
| Cr | -1.15 | 1.99 |
| Ni | -1.38 | 1.90 |
| Fe | -1.53 | 1.94 |
| Ti | -2.96 | 2.04 |
| Mn | -0.67 | 1.97 |

**3.2 Mechanical and dynamical stability**

In the next stage to understand the mechanical stability of the system as given in Table 2, we have studied different elastic coefficients, such as bulk modulus (B), shear modulus (G), Young's modulus (E), Poisson's ratio (ν), and Pugh ratio (B/G). These constants were derived from the second-order derivative matrices, using the relation

Here $C_{ij}$ represents the elastic stiffness coefficients, $V_0$ is the equilibrium volume, E is the total energy, and $\epsilon_i$ and $\epsilon_j$ are the applied strains. These calculations provide critical insights into the mechanical stability and deformation response of the material under different stress conditions, offering a comprehensive understanding of its elastic behavior. To calculate all the mechanical properties, we used online USPEX tools and utilities given by Mazhnik and Oganov [46, 47].

**Table 2** Mechanical stability parameters: bulk modulus (B), shear modulus (G), Young's modulus (Y), Poisson's ratio (ν), and Pugh ratio (B/G) of TM-doped 2DPA

| Elastic Modulus | Pure | Transition Metal | | | | | |
|---|---|---|---|---|---|---|---|
| | | Co | Cr | Ni | Fe | Ti | Mn |
| B | 51.7 | 68.4 | 70.2 | 65.7 | 62.7 | 70.9 | 61.5 |
| G | 24.9 | 34.6 | 36 | 31.8 | 30.8 | 36.4 | 30.5 |
| Y | 64.3 | 88.7 | 92.1 | 82 | 79.5 | 93.2 | 78.5 |
| ν | 0.293 | 0.28 | 0.28 | 0.29 | 0.28 | 0.28 | 0.28 |
| B/G | 0.48 | 0.5 | 0.51 | 0.48 | 0.49 | 0.51 | 0.5 |

The results in Table 2 demonstrate notable changes in the mechanical properties of two-dimensional polyaramid (2DPA) upon doping with transition metals (TMs). The bulk modulus, which reflects resistance to volume compression, increases significantly in TM-doped 2DPA, indicating enhanced hardness and structural stability compared to the pristine material. This suggests that TM doping strengthens the material against external pressure. Likewise, the shear modulus and Young's modulus, representing resistance to shape deformation and stiffness, also rise upon doping, confirming improved mechanical robustness of the 2DPA nanosheets. Among the various dopants, certain TMs yield the highest mechanical resistance, suggesting superior structural integrity under stress. Importantly, the Poisson's ratio remains within the ductile regime for all TM-doped systems, indicating that these modifications do not compromise the material's flexibility. The retained ductility is beneficial for applications in flexible electronics and mechanical systems. The Pugh ratio (bulk modulus to shear modulus), a widely used metric to assess ductility versus brittleness [54], also increases after TM doping. A higher Pugh ratio implies better toughness and reduced brittleness. These enhancements are particularly advantageous for devices requiring both durability and flexibility, such as sensors, wearable electronics, and nanocomposite structures. Therefore, TM doping effectively improves the stability, stiffness, and toughness of 2DPA while preserving its flexible nature, expanding its potential in multifunctional applications, including energy storage and flexible electronics. TM-induced modification of surface properties thus opens pathways for tailoring 2D materials for a range of technological uses.

### 3.3 Phonon Dispersion Study of Pristine 2DPA

The phonon dispersion of pristine 2D polyaramid (2DPA) was computed to evaluate its vibrational characteristics and dynamical stability. The phonon spectrum, plotted along high-symmetry directions of the Brillouin zone, is shown in Figure X. The absence of any imaginary (i.e., negative) frequencies across the entire Brillouin zone confirms the dynamical stability of the 2DPA monolayer. This result validates the structural integrity of the relaxed configuration and indicates that the system resides at a true energy minimum without soft modes that would otherwise drive a structural distortion.

The dispersion spectrum comprises three acoustic branches and a set of optical branches, the number of which corresponds to the vibrational degrees of freedom of the unit cell. The acoustic branches include:

**Out-of-plane acoustic (ZA) mode:** Characterized by a quadratic dispersion near the Γ-point, typical for 2D materials, with frequencies ranging from 0 up to ~50 cm$^{-1}$ (1.50 THz). This mode reflects flexural vibrations perpendicular to the plane and is sensitive to interlayer interactions (if any) and out-of-plane stiffness.

**Transverse acoustic (TA) and Longitudinal acoustic (LA) modes:** These exhibit linear dispersion near the Γ-point and extend up to approximately 100 to 150 cm$^{-1}$ (i.e., 2.998 to 4.497 THz). The LA mode corresponds to in-plane vibrations along the wave vector direction, while the TA mode corresponds to in-plane motion perpendicular to the wave vector. Their relatively steep slopes near Γ suggest high group velocities and thus contribute positively to in-plane thermal conductivity.

The optical phonon modes are distributed above the acoustic region and span a broader frequency range, extending from ~150 cm$^{-1}$ (4.497 THz) to above 300 cm$^{-1}$ (~9.00 THz). These can be further categorized as follows:

**Low- to mid-frequency optical modes (~150–250 cm$^{-1}$):** These modes are likely associated with bending vibrations of the aromatic backbone and deformations involving hydrogen atoms attached to nitrogen or carbon sites.

**High-frequency optical modes (~250–330 cm$^{-1}$):** These are dominated by C–C and C–N bond stretching vibrations within the aromatic polyaramid ring structure. The presence of high-frequency modes at the upper end of the spectrum reflects the strong covalent nature of these bonds and the light mass of the involved atoms.

Furthermore, there is a distinct separation or gap between the acoustic and optical branches, with a relatively sparse density of states in the ~150 cm$^{-1}$ region. This phonon band gap plays a crucial role in suppressing scattering between acoustic and optical phonons, which in turn favors enhanced lattice thermal conductivity and contributes to phonon coherence in thermal transport. The clear resolution of individual branches and the absence of band-crossing or mixing between acoustic and optical modes indicate well-defined phonon modes and vibrational eigenstates. This phonon profile is characteristic of well-ordered, crystalline 2D covalent organic frameworks and reflects the anisotropic bonding topology of polyaramid. The phonon dispersion reveals several key insights: The dynamical stability of the pristine 2DPA system is confirmed. The quadratic ZA branch and steep LA/TA branches imply mechanical flexibility and high in-plane thermal conductivity, respectively. The broad high-frequency optical region points to strong intra-layer bonding. The acoustic-optic phonon gap can contribute to reduced anharmonic scattering and enhanced phonon lifetimes. These characteristics collectively establish 2DPA as a promising material platform for thermally stable, flexible electronics, where both vibrational stability and controlled thermal transport are crucial.

The distinct behavior of acoustic and optical modes underscores the material's robustness and potential for engineering tunable thermal properties [55]. These insights reinforce 2DPA's suitability for advanced applications in electronics, thermal systems, and energy-related technologies [56].

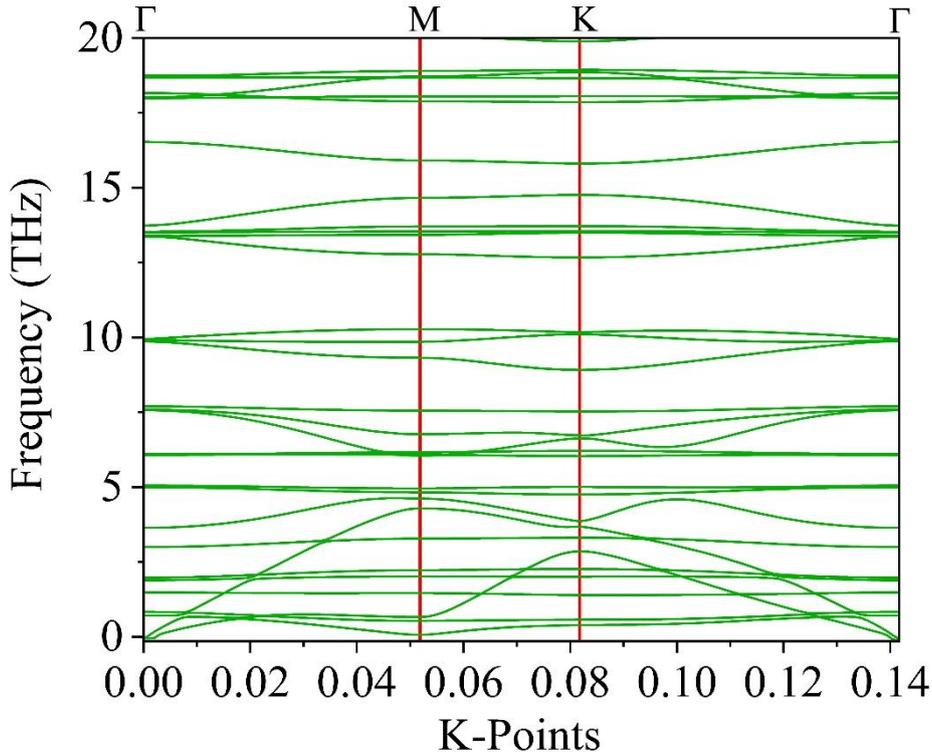
**Figure 2.** Phonon dispersion of pristine 2DPA

### 3.4 Electronic Properties analysis through density of states and band structure

A comprehensive understanding of the electronic and magnetic properties of two-dimensional (2D) materials is essential for their deployment in advanced technologies such as transistors, sensors, and spintronic devices. To explore the electronic characteristics of 2D polyaramid (2DPA), we conducted a detailed analysis of its band structure and partial density of states (DOS) along high-symmetry points (Γ, M, and K) in the Brillouin zone (Figure 3). These results offer critical insights into the band gap, magnetic nature, and overall electronic behavior, enabling further material design. The band structure reveals that pristine 2DPA is a direct band gap semiconductor with a gap of approximately 2.56 eV, where both the conduction band minimum and valence band maximum occur at the Γ point. This value aligns well with previous theoretical predictions [51] and suggests that 2DPA is a strong candidate for optoelectronic applications like transistors and photodetectors. The direct nature of the band gap is particularly favorable for efficient electron transitions and light emission. Analysis of the DOS (Figure 3) further shows that the valence band primarily consists of carbon 2p states, while the conduction band features carbon 2p and nitrogen 2p states, with minor nitrogen 2s contributions. This hybridization enhances stability and contributes to the material's semiconducting character. Both majority and minority spin states are symmetric, confirming the nonmagnetic character of pristine 2DPA. This behavior is typical of nonmagnetic semiconductors and aligns with findings for other 2D materials such as graphene oxide and TMDs [51–53]. Studies on polyaramid-based systems also confirm that their electronic properties are dominated by semiconducting features without intrinsic magnetism [48]. The absence of magnetic ordering suggests potential for magnetic tuning through external modifications. Although pristine 2DPA lacks magnetism, introducing transition metal (TM) dopants such as Co, Fe, and Mn can induce localized magnetic moments and modify its properties. Previous reports confirm that TM doping in 2D materials like graphene and TMDs can result in magnetic ordering, making these systems attractive for spintronic and storage applications [53–58]. In our study, doping 2DPA with Co and Fe introduces significant magnetic moments, indicating the potential emergence of ferromagnetic or half-metallic behavior. These effects are attributed to strong interactions between the dopants and the 2DPA matrix. Hence, while pristine 2DPA shows semiconducting and nonmagnetic characteristics, it can be tailored through TM doping to exhibit desired electronic and magnetic functionalities. This adaptability underscores its potential in both traditional electronics and spintronic technologies. Transition metal doping significantly alters the DOS of 2DPA, influencing its electronic, magnetic, and optical responses. The incorporation of TMs such as Co, Cr, Ni, Fe, Ti, and Mn leads to strong interactions between the dopant atoms and the host matrix, affecting charge distribution and structural properties. These interactions can introduce localized states, reduce band gaps, and induce magnetism, thus broadening the scope for advanced functional devices. To understand these effects, we performed charge density difference (CDD) and band structure analyses (Figures 4 and 5). The CDD plots show notable electron redistribution around dopant sites, particularly at carbon-nitrogen bonds, with pronounced effects for Fe and Co due to their d-electron characteristics. These changes highlight strong dopant-host coupling. Band structure calculations reveal significant changes in the electronic landscape upon TM doping. New states emerge near the Fermi level, shifting its position and modifying the material's band gap. For example, Fe and Co introduce spin-polarized states due to their unpaired d-electrons, resulting in exchange interactions that trigger magnetic ordering within the 2DPA matrix. Such spin polarization is key for spintronic devices that exploit electron spin manipulation. These observations mirror previous findings on other 2D materials, where TM doping induces magnetism and tailors band structures [58, 59]. For 2DPA, this strategy opens new directions for designing materials with specific magnetic and conductive traits. By controlling dopant type and concentration, 2DPA can be engineered for targeted use in magnetoelectronics and energy-related systems. Our results confirm that TM doping transforms 2DPA from a nonmagnetic semiconductor into a material with tunable conductivity and magnetism. Specifically, doping with Ti ($3d^2$, $4s^2$), Cr ($3d^5$, $4s^1$), Fe ($3d^6$, $4s^2$), Co ($3d^7$, $4s^2$), and Ni ($3d^8$, $4s^2$) introduces n-type electronic states and shifts the Fermi level. The resulting band gaps for Ti-, Cr-, Fe-, Co-, and Ni-doped systems are 0.65 eV, 0.61 eV, 0.26 eV, 0.58 eV, and 0.70 eV, respectively, as depicted in Figures 4 and 5. In contrast, Mn ($3d^7$, $4s^2$) doping leads to metallic behavior, evidenced by states at the Fermi level, indicating a transition from semiconducting to metallic character.

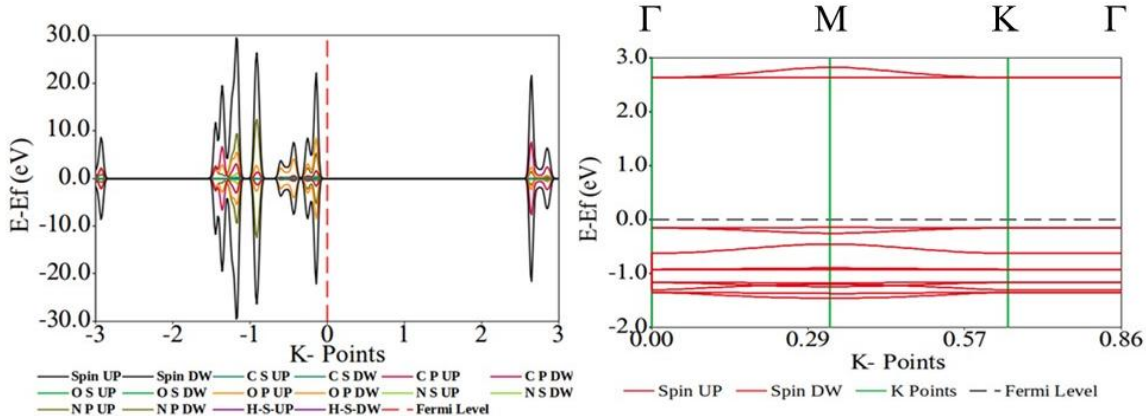

**Figure 3** Electronic densities of states and band structure for the pristine 2DPA exhibiting semiconductor nature within the energy range of 2.56 eV (The dotted line in DOS shows the Fermi energy).

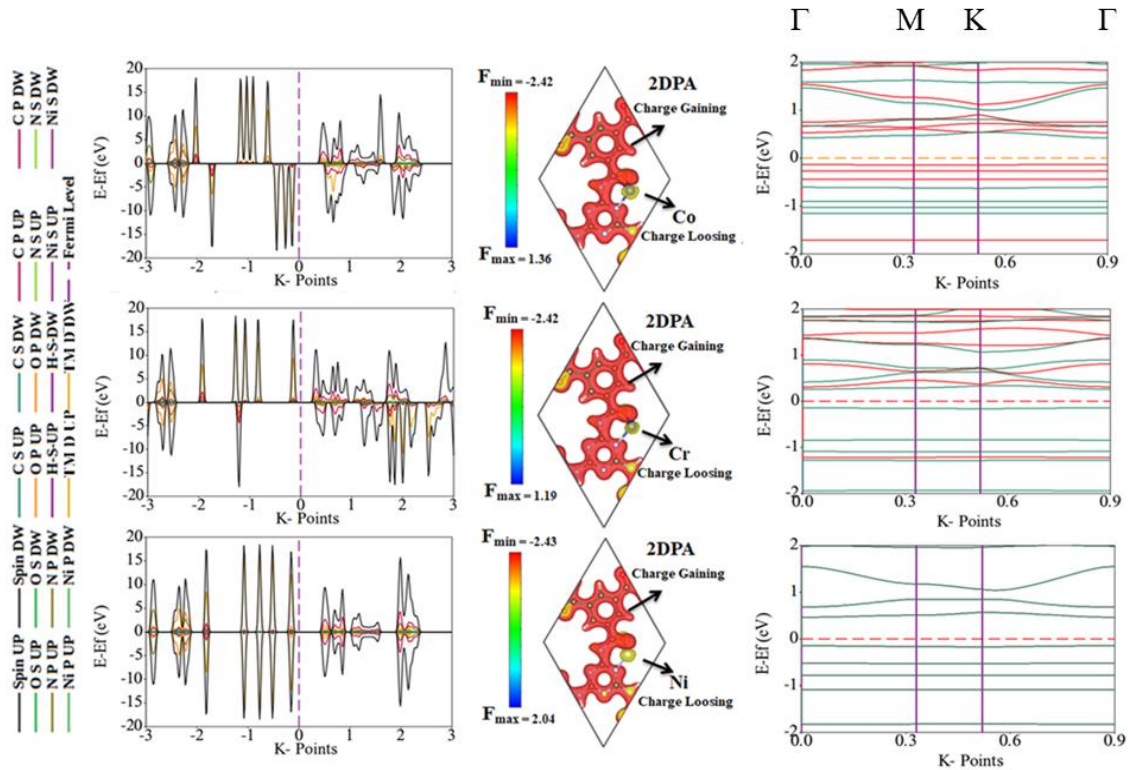

**Figure 4** Electronic density of states, charge density difference (Iso-surface 0.13), and band structure plot for the Co, Cr, and Ni-doped 2DPA (Dotted line in DOS is showing Fermi energy).

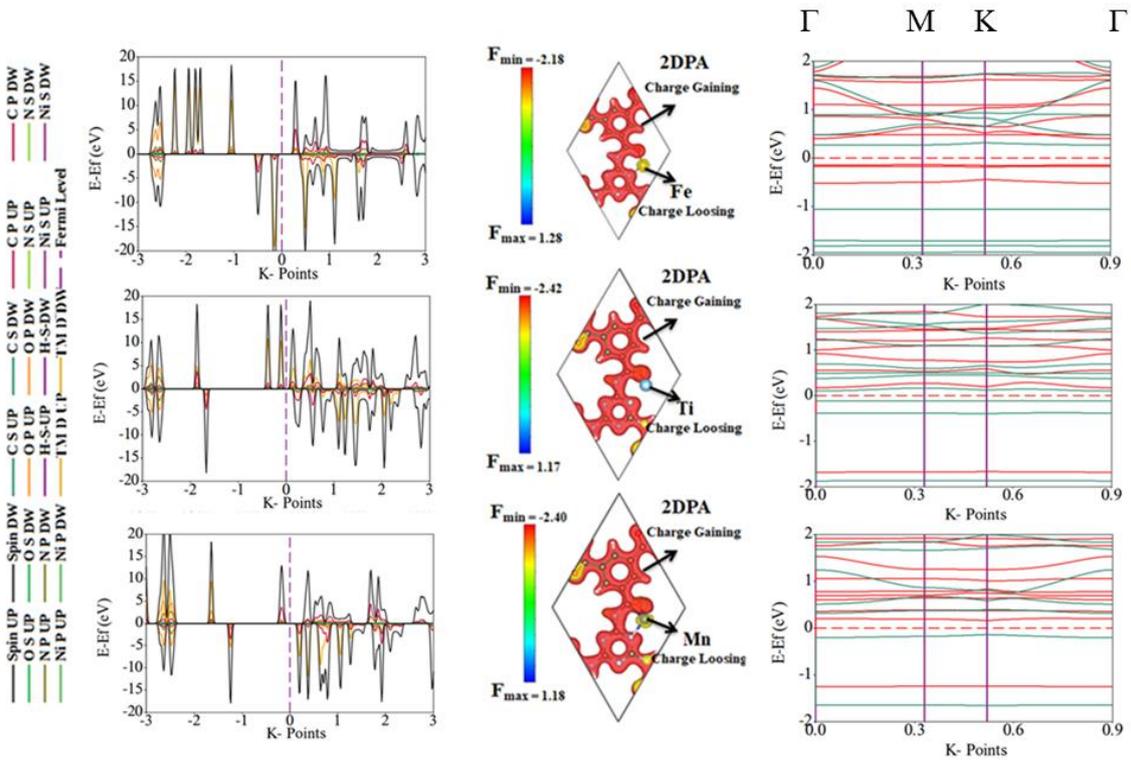

**Figure 5.** Electronic density of states, charge density difference (iso-surface 0.13), and band structure plot for the Fe, Ti, and Mn-doped 2DPA. (The dotted line in DOS shows the Fermi energy.)

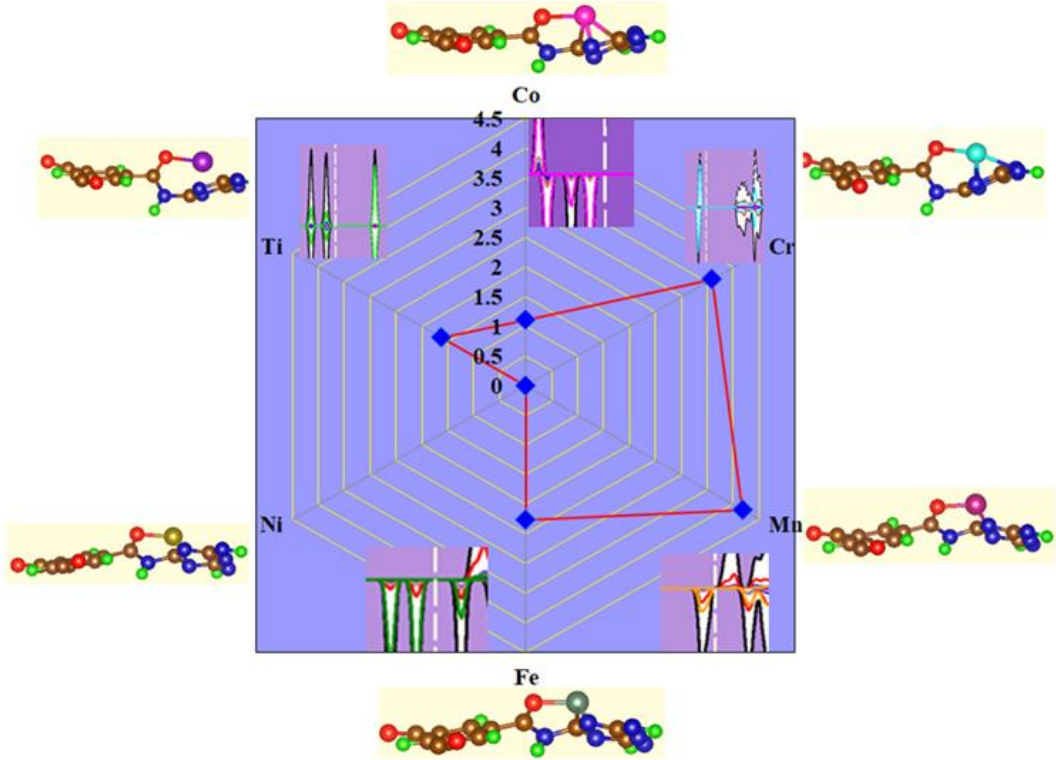

**Figure 6** Schematic representation of induced magnetic moment in Co, Cr, Mn, Fe, and Ti-doped 2DPA in the radar chart with stable configurations (3dyz, 3dxy, 3dz$^2$ and 2pz, 2px orbital contributions of TM atoms and (N, O atoms) 2DPA system.

## 3.5 Charge transfer mechanism

Charge density difference (CDD) plots provide further insight into the effects of transition metal (TM) doping on the electronic structure of 2D polyaramid (2DPA). In these plots, yellow and red regions represent positive and negative charge transfers, respectively, which help visualize how charges are redistributed upon doping. The results indicate that the polarized charges primarily originate from the localized 3d electrons of the transition metal atoms, with only a minor contribution from the surrounding carbon (C), nitrogen (N), and oxygen (O) atoms of the 2DPA matrix. This observation is in agreement with previous studies, such as the work by Tiwari et al. [60], which also showed that transition metal dopants lead to significant charge transfer primarily from the metal atoms. The TM dopants, having lower electronegativity compared to the 2DPA atoms, act as electron donors. In contrast, the higher electronegativity atoms in the 2DPA structure—particularly nitrogen, oxygen, and carbon—function as electron acceptors. This donor-acceptor interaction between the transition metals and the surrounding 2DPA atoms further influences the electronic properties, enhancing the overall electron transfer and polarization within the material.

This results in a net charge transfer from the transition metals to the 2D polyaramid (2DPA) material, which is consistent with the findings in the literature on transition metal (TM) doping in other 2D materials, such as graphene and $MoS_2$. Previous studies have demonstrated that TM doping can effectively modulate the electronic properties of 2D materials by shifting the Fermi level, altering the band gap, and inducing spin-polarization [61-63]. These modifications play a crucial role in optimizing materials for various applications, particularly in nanoelectronics and spintronics, where precise control over the band gap and magnetic properties is essential for device performance. The substantial charge transfer observed in the present study indicates that there is enhanced covalent bonding between the transition metal dopants and the nitrogen (N), oxygen (O), and carbon (C) atoms in the 2DPA system, further contributing to the material's altered electronic structure and potential for advanced technological applications.

## 3.6 Magnetic characteristics

Transition metals (TMs) possess partially filled d-orbitals that can interact with the electronic states of the polyaramid material, especially the π-electrons in its conjugated system. This hybridization between the d-orbitals of transition metals and the π-electrons of the 2D polyaramid (2DPA) can create new electronic states, which significantly modify the material's electronic properties. Transition metals such as Co, Cr, Mn, Fe, Ni, and Ti, all of which have unpaired d-electrons, can induce magnetism in the 2D polyaramid structure. The magnetic behavior arises from the interaction between the magnetic moments of these transition metals and the electronic states of the material, resulting in spin polarization. This leads to an imbalance in the density of states (DOS) for spin-up and spin-down electrons, a phenomenon observed in the DOS calculations for Co, Cr, Mn, Fe, and Ti-doped 2DPA (Figs. 4 and 5). The magnetic ordering in the system is predominantly driven by indirect exchange interactions between the pz orbitals of nitrogen or the px orbitals of oxygen in the 2DPA structure and the dxz/dyz and dz² orbitals of the transition metal dopants. These interactions lead to spin-splitting of the energy bands, which is visible in the projected density of states (PDOS) for each of the doped metals (Fig. 4). For instance, Co, Cr, Mn, Fe, and Ti dopants generate magnetic moments of 1.14 μB, 3.57 μB, 2.26 μB, 4.19 μB, and 1.62 μB per unit cell of the 2DPA surface, respectively (Fig. 6). These values indicate that these metals significantly alter the magnetic properties of 2DPA, with Co, Cr, Mn, and Fe inducing strong magnetic moments in the material. Notably, the magnetic moments are primarily localized around the dopant atoms, with the magnetization being considerably higher at the dopant site than in the surrounding lattice, emphasizing the localized nature of the induced magnetism. However, it is important to note that not all transition metals induce magnetism in the 2DPA system. Specifically, Ni, a first-row transition metal, does not produce any magnetic moment in the 2DPA surface. This lack of magnetism can be attributed to the symmetry behavior of spin-up and spin-down channels in PDOS (Figure 4) between the Ni atom and the surrounding O and C atoms in the 2DPA lattice, which leads to the cancellation of the magnetic moments. Further to verify the ferromagnetic and antiferromagnetic order, the total energy comparison between magnetic configurations (FM & AFM) is tabulated in Table 3.

Table 3: Total energy difference between FM and AFM states of the TM-doped 2DPA system

| System | $E_{FM}$ | $E_{AFM}$ | $\Delta E = E_{AFM} - E_{FM}$ |
|---|---|---|---|
| Co.2DPA | -203.26 | -202.11 | 1.15 |
| Cr.2DPA | -206.68 | -203.55 | 3.13 |
| Mn.2DPA | -204.65 | -203.01 | 1.64 |
| Fe.2DPA | -206.23 | -203.13 | 3.1 |
| Ni.2DPA | -202.01 | -201.36 | 0.65 |
| Ti.2DPA | -204.89 | -203.77 | 1.12 |

As can be seen in Table 3, the difference between $E_{AFM}-E_{FM}$ is high for Co, Cr, Mn, Fe, Ni, and Ti, that is indicates ferromagnetic ordering. This behavior contrasts with the strong ferromagnetic ordering seen with other transition metal dopants. These findings align with similar observations in other 2D materials. For example, Hu et al. [63] predicted a magnetic moment of 5.00 μB per supercell for $MoS_2$ and $MoSe_2$ with a single Cu dopant, while they found that copper doping in $MoTe_2$ did not induce magnetism. This lack of magnetism in $MoTe_2$ was attributed to the delicate balance between crystal-field splitting and spin splitting, which is necessary for the development of magnetic properties [65]. These results suggest that the choice of transition metal and its interaction with the host material are critical factors in determining the magnetic behavior of 2D materials, with potential implications for the design of novel spintronic devices.

## Conclusion

In conclusion, our first-principles density functional theory (DFT) calculations provide a comprehensive investigation into the surface magnetic properties and behavior of transition metal (TM)-doped 2D-polyaramid (2DPA) materials. Structural stability is confirmed via mechanical parameters, and dynamical stability is validated through phonon calculations.

On the electronic front, our calculations show that transition metal doping induces strong formation interactions with the 2DPA lattice, with formation energies ranging from -1.15 eV to -2.96 eV. This interaction stabilizes the doped materials in a planar geometry, with the exception of Mn, which exhibits a weaker interaction and formation energy of -0.67 eV. The variation in formation energies, particularly for Mn, indicates that it might introduce distinct modifications in the surface properties, which could influence the material's magnetic characteristics. The incorporation of transition metals also introduces additional electronic states within the 2DPA material, shifting the Fermi level and altering the electronic structure. Notably, doping with Fe results in a significant reduction in the band gap to 0.26 eV, suggesting a transition to metallic behavior. This reduction of the band gap offers potential for applications in advanced electronics and magnetism-based devices.

Magnetically, the doped systems exhibit clear ferromagnetic ordering, primarily driven by indirect exchange interactions between the transition metal dopants and the 2DPA lattice. This leads to the generation of magnetic moments in the doped systems, with Co, Cr, Fe, Mn, and Ti inducing magnetic moments of 1.14 $\mu_B$, 3.57 $\mu_B$, 2.26 $\mu_B$, 4.19 $\mu_B$, and 1.62 $\mu_B$ per unit cell, respectively. These significant magnetic moments highlight the role of transition metal doping in engineering the magnetic properties of 2DPA.

These findings underscore the promise of transition metal-doped 2DPA materials in the development of advanced spintronic devices, where both electronic and magnetic properties need to be precisely controlled. The ability to tailor the surface magnetic properties of 2DPA through doping with transition metals offers a versatile platform for designing materials with applications in spintronics, sensors, memory devices, and other next-generation technologies.

## Author contributions

Ravi Trivedi: Conceptualization, Data Curation, Data analysis, Writing-original draft.
Chaithanya P. Bhat: VASP and Phonon calculations in complete revision, editing, and checking the final version of the draft.
Shakti S. Ray: Manuscript Draft and Analysis
Debashis Bandyopadhyay: Supervision and VASP calculations, correcting the final draft.

## Acknowledgments

One of the authors, Debashis Bandyopadhyay, is thankful to the Science and Engineering Research Board (SERB), Government of India, for the research funding provided under sanction order CRG/2022/003249, which enabled this work.

## Declaration of Generative AI and AI-assisted technologies in the writing process

To improve the quality of the language and ensure proper citation, the author(s) used Grammarly and Turnitin software while drafting this manuscript. After utilizing these tools, the author(s) reviewed and made any necessary revisions, and they assume complete responsibility for the publication's content.

## Data availability

The raw/processed data required to reproduce these findings cannot be shared at this time, as the data also forms part of an ongoing study on 2DPA materials. Initial structure coordinates have been shared in the SI file. The corresponding author can be contacted if someone wants to request the data from this study.

## Conflicts of interest

The authors declare no conflict of interest.